\documentclass[12pt]{iopart}

\usepackage{iopams}  
\usepackage{graphicx}
\usepackage{xcolor}
\usepackage{color}
\usepackage{hyperref}
\usepackage[nameinlink,capitalize]{cleveref}
\crefname{section}{Sec.}{Secs.}
\crefname{table}{Tab.}{Tabs.}
\crefname{eqnarray}{Eq.}{Eqs.}

\newcommand{\msun}{$M_\odot$ \,}
\newcommand{\red}[1]{{\color{black}#1}} 

\begin{document}
\title[EOS Parameters from GW Inference]{Direct Inference of Nuclear Equation-of-State Parameters from Gravitational-Wave Observations}

\author[0000-0002-7775-5423]{Brendan~T.~Reed \footnote{Corresponding author}}
\address{Theoretical Division, Los Alamos National Laboratory, Los Alamos, NM 87545, USA}
\ead{breed@lanl.gov}

\author[0000-0002-1198-7774]{Cassandra~L.~Armstrong}
\address{Intelligence and Space Research Division, Los Alamos National Laboratory, Los Alamos, NM 87545, USA}
\ead{clarmstrong@lanl.gov}

\author[0000-0003-0427-3893]{Rahul~Somasundaram}
\address{Theoretical Division, Los Alamos National Laboratory, Los Alamos, NM 87545, USA}
\ead{rsomasundaram@lanl.gov}

\author[0000-0002-9180-5765]{Duncan~A.~Brown}
\address{Department of Physics, Syracuse University, Syracuse, NY 13244, USA}
\ead{dabrown@syr.edu}

\author[0000-0002-0355-5998]{Collin~Capano}
\address{Department of Physics, Syracuse University, Syracuse, NY 13244, USA}
\address{Physics Department, University of Massachusetts Dartmouth, North Dartmouth, MA 02747, USA}

\author[0000-0002-3316-5149]{Soumi~De}
\address{Theoretical Division, Los Alamos National Laboratory, Los Alamos, NM 87545, USA}

\author[0000-0003-2656-6355]{Ingo~Tews}
\address{Theoretical Division, Los Alamos National Laboratory, Los Alamos, NM 87545, USA}
\ead{itews@lanl.gov}
\vspace{10pt}
\begin{indented}
\item[]\date{\today}
\item[]LA-UR-25-25425
\end{indented}
\newpage
\begin{abstract}
The observation of neutron star mergers with gravitational waves (GWs) has provided a new method to constrain the dense-matter equation of state (EOS) and to better understand its nuclear physics. 
However, inferring nuclear microphysics from GW observations necessitates the sampling of EOS model parameters that serve as input for each EOS used during the GW data analysis.
The sampling of the EOS parameters requires solving the Tolman-Oppenheimer-Volkoff (TOV) equations a large number of times -- a process that slows down each likelihood evaluation in the analysis on the order of a few seconds. 
Here, we employ emulators for the TOV equations built using multilayer perceptron neural networks to enable direct inference of nuclear EOS parameters from GW strain data.
Our emulators allow us to rapidly solve the TOV equations, taking in EOS parameters and outputting the associated tidal deformability of a neutron star in only a few tens of milliseconds. 
We implement these emulators in \texttt{PyCBC} to directly infer the EOS parameters using the event GW170817, providing posteriors on these parameters informed solely by GWs.
We benchmark these runs against analyses performed using the full TOV solver and find that the emulators achieve speed ups of nearly \emph{two orders of magnitude}, with negligible differences in the recovered posteriors.
Additionally, we constrain the slope and curvature of the symmetry energy at the 90\% upper credible interval to be $L_{\rm sym}\lesssim106$ MeV and $K_{\rm sym}\lesssim26$ MeV.
\end{abstract}

%
%
%
%
%

\section{Introduction}
In 2017, the LIGO and Virgo gravitational-wave (GW) interferometers detected the collision of two neutron stars (NS) for the first time -- an event labeled GW170817 \cite{LIGOScientific:2017ync,LIGOScientific:2017vwq}.  
This event is the most informative binary NS merger detected to date, and has provided invaluable constraints for the nuclear equation of state (EOS)~\cite{Annala:2017llu,Margalit:2017dij,Radice:2017lry,De:2018uhw,Tews:2018iwm,Greif:2018njt,Most:2018hfd,Capano:2019eae,Dietrich:2020efo,Landry:2020vaw,Reed:2020,Raaijmakers:2021,Miller:2021qha,Han_2021,Somasundaram:2024ykk,Rutherford:2024srk}. 
The primary quantity containing information on the EOS measured by GW data is the tidal deformability (or polarizability) $\Lambda$, which, in its dimensionless form, is given by
\begin{eqnarray}
    \Lambda = \frac{2}{3}k_2\Bigg(\frac{Rc^2}{GM}\Bigg)^5\,.
\end{eqnarray}
Here, $M$ and $R$ are mass and radius of the NS and $k_2$ is the second tidal Love number~\cite{Hinderer_2008,Hinderer_2009}.
The tidal deformability of neutron stars is strongly correlated with the NS radius and is sensitive to the neutron-rich sector of the nuclear EOS. 
Observationally, one can extract a mass-weighted average of the $\Lambda$ of each of the two merging neutron stars, $\tilde\Lambda$, which is given by
\begin{equation}
    \tilde\Lambda = \frac{16}{13} \frac{(12q + 1)\Lambda_1 + (12 + q)q^4 \Lambda_2}{(1 + q)^5}\,.
    \label{eq:ltilde}
\end{equation}
\red{This term first appears in the gravitational waveform at fifth post-Newtonian order.}
Together with constraints on the ratio of the two NS masses, $q=m_1/m_2$, with $m_1\geq m_2$  and the chirp mass,
\begin{equation}
    \mathcal{M}_c = \frac{(m_1 m_2)^{3/5}}{(m_1 + m_2)^{1/5}}\,,
    \label{eq:chirp}
\end{equation}
the measurement of $\tilde{\Lambda}$ from GW strain data allows us to place constraints on the EOS.

Extracting EOS information from GW observations has been extensively studied in the literature in the context of both present~\cite{Annala:2017llu,LIGOScientific:2018cki,Malik:2018zcf,De:2018uhw,Capano:2019eae,Essick:2019ldf,Dietrich:2020efo,Kunert:2021hgm} and next-generation~\cite{Landry:2022rxu,Prakash:2023afe,Iacovelli:2023nbv,Rose:2023uui,Bandopadhyay:2024zrr,Walker:2024loo} GW detectors. 
For example, the original constraint on the tidal deformability from GW170817 was $\Lambda<800$~\cite{LIGOScientific:2017vwq}. 
This constraint is stringent enough to rule out a number of EOS~\cite{Reed:2021nqk} as it requires the EOS to be soft, i.e., to have lower pressure than predicted in these models.
Subsequent analyses of GW170817 have either considered only GW data while neglecting additional information from nuclear physics~\cite{De:2018uhw,Radice:2018ozg,Abbott:2018wiz}, or employed a pregenerated set of EOS to analyze GW data~\cite{Capano:2019eae,Dietrich:2020efo}. 

However, these analyses have only constrained the overall EOS in the pressure-density space. 
In contrast, the nuclear microphysics is encoded in a set of EOS parameters. Extracting information on these parameters from GW inferences of the EOS is difficult and may lead to systematic uncertainties; for example, due to limited resolution of pregenerated EOS sets~\cite{Rose:2023uui}.
Furthermore, the EOS parameters might exhibit degeneracies that are difficult to disentangle after an inference run, and implementing different Bayesian priors on these parameters is difficult {\it a posteriori}.
To understand and alleviate some of these concerns, we would like to \emph{directly} sample the microphysical parameters that describe the EOS in analyses of GW data.

Direct sampling of the EOS model parameters during a full GW inference has, thus far, been accomplished only at significant computational expense~\cite{LIGOScientific:2018cki}.
To achieve this, one needs to generate a NS EOS in beta equilibrium from a set of EOS model parameters and solve the Tolman-Oppenheimer-Volkoff (TOV) equations to extract the NS structure on-the-fly.
Depending on the approach, the first step can be completed quickly ($<<1s$)~\cite{Margueron:2017lup}, sometimes employing surrogate modeling~\cite{Armstrong:2025tza}, whereas the second step can take a few seconds on a single CPU. 
In recent years, however, machine-learning (ML) advancements have allowed for large speedups in calculations in several areas of science, including astrophysics~\cite{Pang:2022rzc,giuliani2023bayes,reed:2024,Somasundaram:2024zse,Dax:2024mcn,Desai:2024hlp,Armstrong:2025tza,King:2025tqo}.
In this paper, we use ML to develop tools that allow us to constrain nuclear EOS model parameters directly from the analysis of GW strain data.
In a previous work, we have developed emulators for the TOV equations which generate the tidal deformability $\Lambda$ for EOS models parameterized by different sets of parameters~\cite{reed:2024}.
Our emulators have achieved uncertainties of less than $0.1$\% with evaluation times of a microsecond. 
Here, we implement these emulators in the \texttt{PyCBC} computational framework
\cite{Biwer:2018osg,alex_nitz_2024_10473621} to perform fast and accurate inference of EOS model parameters directly from the strain data of GW170817, provide direct posteriors of these parameters, and assess the systematic uncertainties of such analyses.
\red{In future work, we will further develop this approach by, e.g., using more flexible EOS models.  
While this study mainly focuses on GW170817, additional observations can easily be adapted into our Bayesian framework we present below.}

This paper is structured as follows. 
We summarize the EOS model in \cref{subsec:meta_model} and detail the framework in which the emulators are constructed in \cref{subsec:emulator}.
Our methodology of GW inference and details of the EOS parameter sampling are discussed in \cref{subsec:pycbc}. 
We present and discuss our emulator results, including an assessment of statistical and systematic uncertainties, in \cref{subsec:statistics} and the posteriors obtained with both 5- and 10-parameter EOS models in \cref{subsec:posteriors}. 
We then discuss the implications of our results in \cref{sec:discussion} before concluding in \cref{sec:conclusion}.

\section{Methods}

\subsection{Equation of State Model}
\label{subsec:meta_model}

For generating the EOS, we use the method described in Ref.~\cite{reed:2024} which we summarize here. 
We use a fixed crust EOS model, taken from Ref.~\cite{Douchin:2001}, that is attached to the EOS of the core following the procedure outlined in Ref.~\cite{Koehn:2024set}.
For the EOS of the NS core, at densities below twice nuclear saturation density ($n_0$), we employ the metamodel of Refs.~\cite{Margueron:2017eqc} and~\cite{Margueron:2017lup}. 
The metamodel describes the EOS as a function of density and proton fraction in terms of nucleonic degrees of freedom using a parameterization in terms of the bulk nuclear EOS parameters~\cite{Margueron:2017eqc}.
The metamodel uses the common expansion of the nuclear matter EOS in powers of the proton-neutron asymmetry $\alpha=\frac{n_n-n_p}{n}$,
\begin{equation}
    E(n,\alpha)=\mathcal{E}(n)+\alpha^2 S_2(n)+\mathcal{O}(\alpha^4)\,,
\end{equation}
where $E(n,\alpha)$ is the total energy per particle, $\mathcal{E}(n)$ is the energy of symmetric nuclear matter, and $S_2(n)$ is the quadratic approximation to the symmetry energy~\cite{Somasundaram:2020chb}. 
The total symmetry energy $S(n)$ is defined as the difference between the energy of pure neutron matter and symmetric nuclear matter, i.e. $S(n) \equiv E(n,\alpha=1)-\mathcal{E}(n)$.  
The bulk nuclear EOS parameters are defined via Taylor expansions of symmetric nuclear matter and the symmetry energy around nuclear saturation density,
\begin{eqnarray}
    &&\mathcal{E}(n)=E_{\rm sat}+\frac{1}{2}K_{\rm sat}x^2+\frac{1}{6}Q_{\rm sat}x^3+\frac{1}{24}Z_{\rm sat}x^4 + \dots \,,\\
    &&S(n)=E_{\rm sym}+L_{\rm sym}x+\frac{1}{2}K_{\rm sym}x^2+\frac{1}{6}Q_{\rm sym}x^3 +\frac{1}{24}Z_{\rm sym}x^4 + \dots\,,
\end{eqnarray}
where $x=\frac{n-n_0}{3n_0}$.
By suitable adjustment of the EOS parameters, the metamodel is capable of reproducing the EOS of a large number of nucleonic models. 

At higher densities, the assumption that the EOS can be described by nucleonic degrees of freedom may break down. 
To account for this possibility, we employ a speed-of-sound model to describe the EOS at densities above $2 n_0$ \cite{Tews:2018kmu,Greif:2018njt,Koehn:2024set}.
The parameters of this model are the values of the squared sound speed at discrete density points, the number of which depends on the level of complexity one wishes to employ. 
We connect each low-density EOS, described by the metamodel and crust, with a high-density speed-of-sound model that can be integrated to obtain the pressure, energy density, and chemical potential at all densities.
In this way, we robustly account for the uncertainty at high densities without making strong model assumptions.

In Ref.~\cite{reed:2024}, we have generated three EOS sets using the previous prescription. 
The three sets vary in the number of EOS parameters and speeds of sound employed in each set (1, 5, and 10 parameters in total).
Constraints from microscopic ab-initio nuclear theory, e.g., from chiral effective field theory (EFT)~\cite{Hebeler:2013nza,Tews:2018kmu,Drischler:2017wtt,Keller:2022crb}, and nuclear experiments, such as the lead radius experiment (PREX)~\cite{2021PhRvL.126q2502A,Reed:2021nqk}, can in principle be incorporated by choice of suitable priors on the EOS parameters~\cite{Koehn:2024set}.
\red{Our priors on the nuclear parameters encompass constraints extracted from both experimental and theoretical approaches \cite{Lattimer:2000nx,Reed:2024c,Tews:2018kmu} and our list of prior bounds is given in \cref{tab:combined-priors}. 
We then generate several NS EOSs based on these priors.}
In this paper, we only use the 5- and 10-parameter models as the 1-parameter model lacks reliability and complexity in producing realistic EOSs~\cite{reed:2024}.

The 5-parameter and 10-parameter models both vary K$_{\rm sat}$, L$_{\rm sym}$, and K$_{\rm sym}$ while fixing $E_{\rm sat}=-16$~MeV, $n_0=0.16$~fm$^{-3}$, and $E_{\rm sym}=32$~MeV.
\red{These fixed values are consistent with nuclear structure as determined from nuclear experiments \cite{Lattimer:2004pg,Roca-Maza:2015eza,Drischler:2017wtt,Reed:2024c} and do not considerably impact NS structure.}
In both models, all other nuclear parameters are set to zero.
At $2n_0$, these models transition to the sound speed models, with the 5-parameter model parameterizing the sound speeds at $3n_0$ and $5n_0$ and the 10-parameter model parameterizing the sound speeds in intervals of $n_0$ between $3n_0$ and $9n_0$.
These points are then connected by \red{linear interpolation functions, ensuring continuity in the EOS at both interfaces}.
The parameter vector for both EOS sets, and subsequently the emulators, is then given by
\begin{equation}
    \vec{\theta} = \{K_{\rm sat},L_{\rm sym},K_{\rm sym},c_s^2(n_i)\}\,,
\end{equation}
where the $n_i$ are integer multiples of $n_0$ corresponding to the transition points in either model.

In summary, our EOS sets are constructed using the metamodel which defines the EOS of the core below 0.32 fm$^{-3}$\red{, including electron degrees of freedom to ensure charge and beta equilibrium are achieved}.
The crust EOS of Ref.~\cite{Douchin:2001} is used up to the crust-core transition density of this crust model \red{$\approx 0.076$fm$^{-1}$}and then attached to the metamodel EOS~\cite{Koehn:2024set}.
\red{We note that the crust-core transition density depends on the choice of NEPs and our choice of a fixed crust-core transition density might introduce a slight bias in our analysis.
We will study EOS with self-consistent crust-core transition densities in future work.}
Above these densities we utilize a sound-speed model for the EOS defined in terms of either two or seven squared sound-speed values. 
These EOSs are finally used as input to solve the TOV equations and the corresponding equations for the tidal love number~\cite{Hinderer_2008} resulting in the masses, radii, and tidal deformabilities of the NS~\cite{Tews:2018kmu,Somasundaram:2021clp}.

\subsection{Construction of Emulators}
\label{subsec:emulator}
\red{The emulators we employ here were developed in Ref.~\cite{reed:2024}.}
Using our two EOS sets, specified by their respective model parameters and the mass-tidal deformability relations, we built two emulators for $\Lambda(M;\vec{\theta})$ using multilayer perceptrons (MLP), i.e. feed-forward neural networks~\cite{rosenblatt1958perceptron,MLP}. 
The MLPs are constructed using 5 hidden layers, each containing 64 neurons with the rectified linear unit chosen for the activation function. 
The output layer uses the identity function for the neurons. 
To improve accuracy, we use a bagging ensemble~\cite{Bagging} of $100$ identical MLPs trained independently on the same data. 
We construct an MLP emulator that takes the parameters $\vec{\theta}$ as input and predicts the values of $\log_{10}({\Lambda(M;\vec{\theta}}))$ on a mass grid with $30$ points uniformly spaced between $1M_{\odot}$ and $2M_{\odot}$ for both the 5-parameter and 10-parameter sets. 
The results are then interpolated in $M$ using a cubic spline before being exponentiated back to $\Lambda(M;\vec{\theta})$ as the final output.
In training the emulators, we use 100,000 training samples and validate using another 100,000 samples.
Our emulators reproduced the results for the full TOV solver with an average uncertainty of $0.1\%$ for the tidal deformability of a $1.4M_{\odot}$ NS, $\Lambda_{1.4}$, in a fraction of the time~\cite{reed:2024}.

In the following, we will use nested sampling to sample the EOS parameters directly during data analyses.
During the sampling process, some combinations of EOS parameters result in a maximum NS mass of less than 2$M_\odot$ -- a space we do not train our emulators on. 
To handle this problem, we also built machine-learning classifiers for each of our EOS sets that identify whether a parameter set produces a maximum mass outside the training range.
For both of our classifiers, we use a C-support vector classification (SVC)~\cite{Crammer:2002} with a radial basis function kernel~\cite{scikit-learn}. 
We train our classifiers using 90,000 training samples and validate using 10,000 samples.
The output is a binary (0,1) determining whether the input EOS parameters produce a maximum mass of at least $2M_\odot$ or not. 
For each of our SVC classifiers we find an accuracy of over 99.99\%.

\red{We emphasize that the construction and implementation of our emulators only depends on the choice of the EOS model.
Thus, one can use these emulators for inferences of \textit{any} GW event from a binary neutron star merger, if the two constituent masses are within 1-2\msun. 
Combined constraints on the nuclear parameters could then be obtained by taking the product of likelihood values over multiple events, since the EOS of nuclear matter is expected to be the same for all observed events.
The particular choice of EOS model introduces some bias in our results, but our EOS model is very generic and the framework flexible enough to use  different EOS models in the future.}

\subsection{\texttt{PyCBC} Inference}
\label{subsec:pycbc}

\begin{table}[tb]
    \centering
    \begin{tabular}{| c | c |}
    \hline
    \multicolumn{2}{|c|}{\textbf{EOS Priors}} \\
    \hline
    Parameter & Prior Range \\
    \hline
    $c_s^2(n_i)$ & [$10^{-5}$, 1] \\
    $K_{\rm sat}$ & [200, 300] MeV \\
    $L_{\rm sym}$ & [20, 150] MeV \\
    $K_{\rm sym}$ & [-300, 100] MeV \\
    \hline
    \multicolumn{2}{|c|}{\textbf{GW Priors}} \\
    \hline
    Parameter & Prior Range \\
    \hline
    $\mathcal{M}_c^{\rm src}$ & [1.185, 1.1875) $M_\odot$\\
    $q$ & [1, 2) \\
    $\Delta t_c$ & [-0.1, 0.1) s\\
    $\cos\iota$ & [-1, 1) \\
    $\chi_{1,2}$ & [-0.05, 0.05] Deg.\\
    \hline
    \end{tabular}
    \caption{Priors used in the generation of EOSs and uniform priors on GW parameters used in the Bayesian inference study.}
    \label{tab:combined-priors}
\end{table}

We perform Bayesian inference on the gravitational-wave signal GW170817 using the \texttt{PyCBC}~\cite{Biwer:2018osg} GW analysis package. 
The posterior distribution functions $p(\vec\theta|d(t),H)$ are calculated from parameters $\vec\theta$ for GW waveform model $H$ and compared to data $d(t)$ from the LIGO Hanford, LIGO Livingston, and Virgo detectors~\cite{Vallisneri_2015,LIGOScientific:2019lzm}
\begin{equation}
    p(\vec\theta|d(t),H) = \frac{p(\vec\theta|H)p(d(t)|\vec\theta,H)}{p(d(t)|H)}\,.
\end{equation}
The sampling parameters $\vec\theta$, some of which serve as input parameters for our emulators, now also serve as input to the GW waveform model $H$, which uses the template SEOBNRv4T\_surrogate~\cite{pratten_2025_14999310}. 
We fix the sky parameters and distance parameters to GW170817~\cite{Soares-Santos_2017,Cantiello_2018} and sample the remaining GW waveform parameters.
These remaining parameters are the source frame chirp mass $\mathcal{M}_c^{\rm src}$, mass ratio $q$, offset of trigger time $\Delta t_c$, inclination angle $\iota$, and the two NS spins $\chi_{1z,2z}$.
We marginalize the polarization angle and phase using a uniform prior from $[0,2\pi)$ for both.
Our complete set of sampling parameters is therefore
\begin{equation}
    \vec\theta = \{\mathcal{M}_c^{\rm src},\,q,\,\Delta t_c,\,\iota,\,\chi_{1z},\,\chi_{2z},\,K_{\rm sat},\,L_{\rm sym},\,K_{\rm sym},\,c_s(n_i)\}\,.
\end{equation}
As priors on the GW parameters, we employ uniform distributions informed from previous GW inference studies~\cite{De:2018uhw}.
The chirp mass $\mathcal{M}_c$ is the best constrained parameter from GW170817~\cite{LIGOScientific:2017vwq} so we choose strict priors of $\sim0.2$\% around the $\mathcal{M}_c$ mean. 
The mass ratio can have a large impact on the results \cite{agathos:2015,De:2018uhw} so we choose a broad prior, allowing ratios from 1 to 2.
For the EOS parameters, we use the same priors that were employed when generating the EOS sets. 
We provide a full table of GW and EOS parameter priors in \cref{tab:combined-priors}.

To determine the posterior distribution of the parameters of GW170817, we use \texttt{PyCBC Inference}~\cite{Biwer:2018osg} with the \texttt{dynesty}~\cite{dynesty} nested sampler. 
We use the relative binning~\cite{Dai:2018} model for fast evaluation of the likelihood. 
To use the emulators as well as the full high-fidelity TOV solver, we separately implement a ``plug-in waveform'' model for \texttt{PyCBC} \cite{plugin}. 
This model takes as input the nuclear parameters, uses the emulator or full solver to generate the $\Lambda$'s for each component object, and then passes these values along with the rest of the parameters to the chosen waveform approximant (in this case, \texttt{SEOBNRv4T\_surrogate}~\cite{Lackey:2018zvw}) to generate the GW.

\section{Results}

We performed our calculations on the National Energy Research Scientific Computing (NERSC) Center's Perlmutter cluster \cite{perlmutterNERSC}. 
Our calculations for all solvers use two nodes with 128 CPUs per node. 
Within \texttt{PyCBC}, we performed cross node computations using MPI~\cite{mpiForum}. 
Below, we will detail the results of the Bayesian inference and detail the sources of possible uncertainty.

\subsection{Sampling Convergence and Emulator Efficiency}
\label{subsec:statistics}

\begin{figure*}
    \centering
    \includegraphics[width=\linewidth]{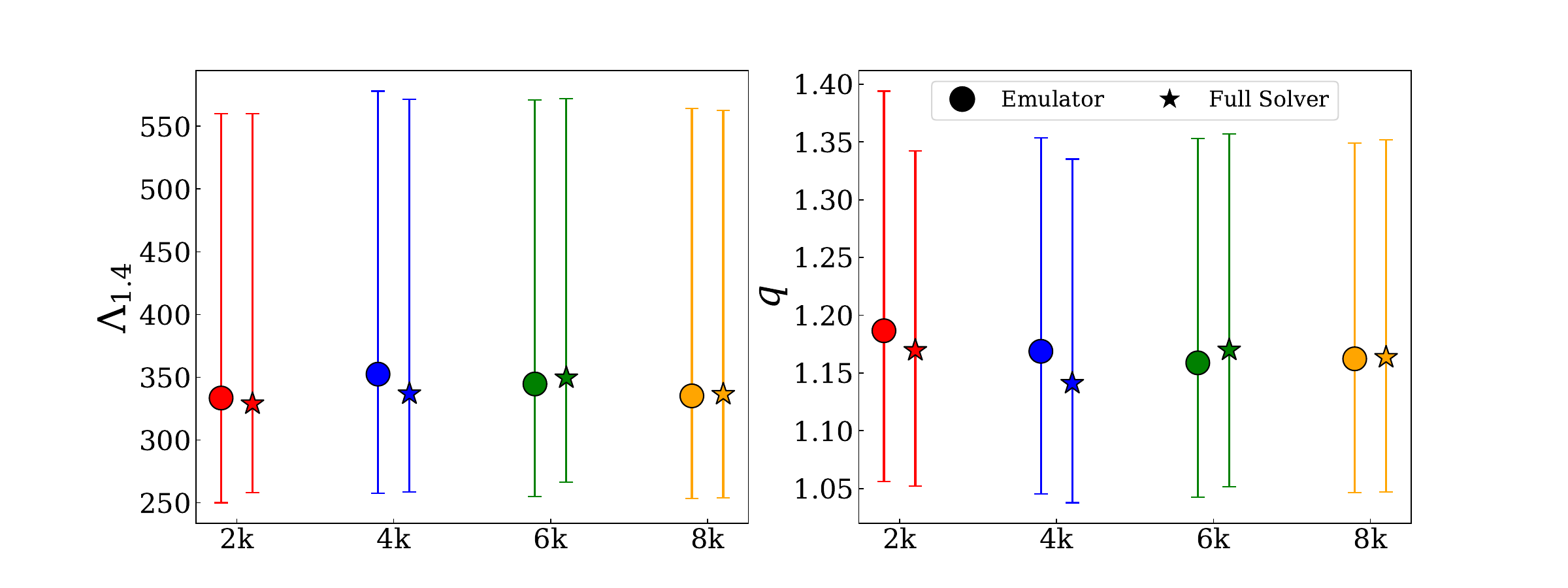}
    \caption{Comparison of convergence for different numbers of live sampling points between the 5-parameter emulator (circle) and the full TOV solver (star). 
    For each setting, the error bars correspond to the 68\% credible interval. 
    Left: Combined tidal deformability $\tilde{\Lambda}$ vs. number of live points. Right: Mass ratio $q$ vs. number of live points. 
    In both cases, we see that the sampling converges for 8000 live points.}
    \label{fig:sampling}
\end{figure*}

\begin{table*}
\centering
\begin{tabular}{| l | c | c |}
\hline
\multicolumn{3}{|c|}{\textbf{GW Parameters}} \\
\hline
Parameter & Emulator & Full Solver \\
\hline
$\mathcal{M}_c^{\rm src}$ ($M_\odot$) & $1.186714_{-0.000092}^{+0.000113}$ & $1.186714_{-0.000092}^{+0.000116}$ \\
$q$ & $1.16_{-0.15}^{+0.33}$ & $1.16_{-0.15}^{+0.33}$ \\
$\Delta t_c$ (s) & $0.00716_{-0.00182}^{+0.00054}$ & $0.00715_{-0.00185}^{+0.00057}$ \\
$\iota$ (Deg.) & $2.566_{-0.085}^{+0.092}$ & $2.566_{-0.084}^{+0.092}$ \\
$\chi_{1z}$ & $0.010_{-0.046}^{+0.035}$ & $0.012_{-0.047}^{+0.033}$ \\
$\chi_{2z}$ & $0.005_{-0.046}^{+0.040}$ & $0.003_{-0.046}^{+0.042}$ \\
\hline
\multicolumn{3}{|c|}{\textbf{EOS Parameters}} \\
\hline
Parameter & Emulator & Full Solver \\
\hline
$K_{\rm sat}$ (MeV) & $238.10_{-34.70}^{+53.50}$ & $239.60_{-36.18}^{+52.65}$ \\
$L_{\rm sym}$ (MeV) & $46.88_{-24.57}^{+62.12}$ & $45.30_{-23.36}^{+63.62}$ \\
$K_{\rm sym}$ (MeV) & $-209.3_{-83.1}^{+228.5}$ & $-214.7_{-78.9}^{+240.8}$ \\
$c_s^2(3n_0)/c^2$ & $0.65_{-0.32}^{+0.31}$ & $0.68_{-0.34}^{+0.29}$ \\
$c_s^2(5n_0)/c^2$ & $0.62_{-0.55}^{+0.35}$ & $0.59_{-0.52}^{+0.38}$ \\
\hline
\end{tabular}
\caption{Comparison of the inferred parameters of GW170817 for the emulator and the full TOV solver using the 5-parameter EOS with 8000 live points. 
We state results at the 90\% credible level.}
\label{tab:gw_corner}
\end{table*}

When performing the inference, there are two main sources of uncertainty that could contribute to a disagreement between the emulator and full solvers: errors resulting from insufficient sampling or errors from the emulator accuracy. 
To address this, we study the convergence of our inference runs with \texttt{PyCBC} with the number of live points for both the high-fidelity (full) solver and the emulator.
In \cref{fig:sampling}, we compare the convergence of the 5-parameter emulator and full TOV solver by comparing the 68\% credible intervals for the $\Lambda_{1.4}$ (left) and mass ratio $q$ (right) posteriors with increasing number of live points.
We calculate $\Lambda_{1.4}$ using our emulators for $\Lambda(m;\vec{\theta})$ but with EOS parameters inferred with either the full solver or the emulator.
The values of $q$ are calculated using the sampled posteriors.
By gradually increasing the number of live points, we find that the central values and 68\% credible intervals converge after 8000 live points for both quantities, showing excellent agreement between emulator and full solver. 

We further support this conclusion by comparing inferred GW and EOS parameters for the 5-parameter model between the emulator and full solver runs with 8000 live points in \cref{tab:gw_corner}.
For these parameters, we obtain nearly identical results for the 90\% credible interval.
Our results are also consistent with the original inference of GW170817 published by the LIGO collaboration~\cite{LIGOScientific:2017vwq}, further indicating that our inference is robust and our approach is suitable for analyzing future GW data from binary NS mergers. 

\begin{table}[htb]
    \centering
    \begin{tabular}{| c | c |}\hline\hline
        \textbf{Deformability Solver} & \textbf{\#CPU Hours} \\\hline
        5-Parameter Emulator & 40\\\hline
        5-Parameter Full Solver & 3100\\\hline
        10-Parameter Emulator & 43 \\\hline
    \end{tabular}
    \caption{Computational time per \texttt{PyCBC} run for different EOS models and emulators vs. full solver in CPU hours.}
    \label{tab:cpu_hours}
\end{table}

While both posteriors are consistent with each other, there is a small difference which we attribute to emulator errors.
By calculating the weighted average difference between the 90\% credible intervals for the EOS parameters for both the emulator and the full solver, we find an average weighted mean difference of $\approx4\%$.
The largest contribution to this difference stems from $c_s^2(3n_0)$, where full solver and emulator disagree by approximately $5\%$.
The smallest contribution stems from $K_{\rm sat}$, with an approximate difference of $0.6\%$.
The bulk of these discrepancies likely arise from \red{several calls to the emulators, whose errors, although small compared to the uncertainties in the waveform itself, accumulate over time}.
However, we conclude that with 8000 live points these differences are negligible compared to uncertainties of the EOS parameters themselves.
\red{Thus, we do not explicitly include emulator errors in the subsequent analysis.}

While we sacrifice some accuracy by employing emulators, we gain a substantial speed up of approximately two orders of magnitude over using the full solver\footnote{The speed-ups vary slightly for different numbers of live points, ranging from 75-80 times.}, see \cref{tab:cpu_hours}.
This is consistent with the speed-ups of singular function calls for the MLP emulator as reported in Ref.~\cite{reed:2024}: 0.028s for the emulator and 2.4s for the full TOV solver on average. 
We also find that within the 10-parameter model, we only experience a slight slowdown of the MLP framework despite doubling the number of EOS sampling parameters. 
We conclude that the emulated runs provide a fast way to perform detailed GW inference while maintaining high accuracy.
\red{We did not perform a comparison between the full-solver and the emulator runs for the 10-parameter model due to its computational expense. 
Comparing the high-fidelity and emulator results for the 5-parameter model showed excellent agreement between the two.
Because the emulator errors for the 5- and 10-parameter models are very similar~\cite{reed:2024}, we do not expect our conclusions to change in the latter case.
}

\subsection{GW Inference Posteriors}
\label{subsec:posteriors}

\begin{figure}
    \centering
    \includegraphics[width=1\linewidth]{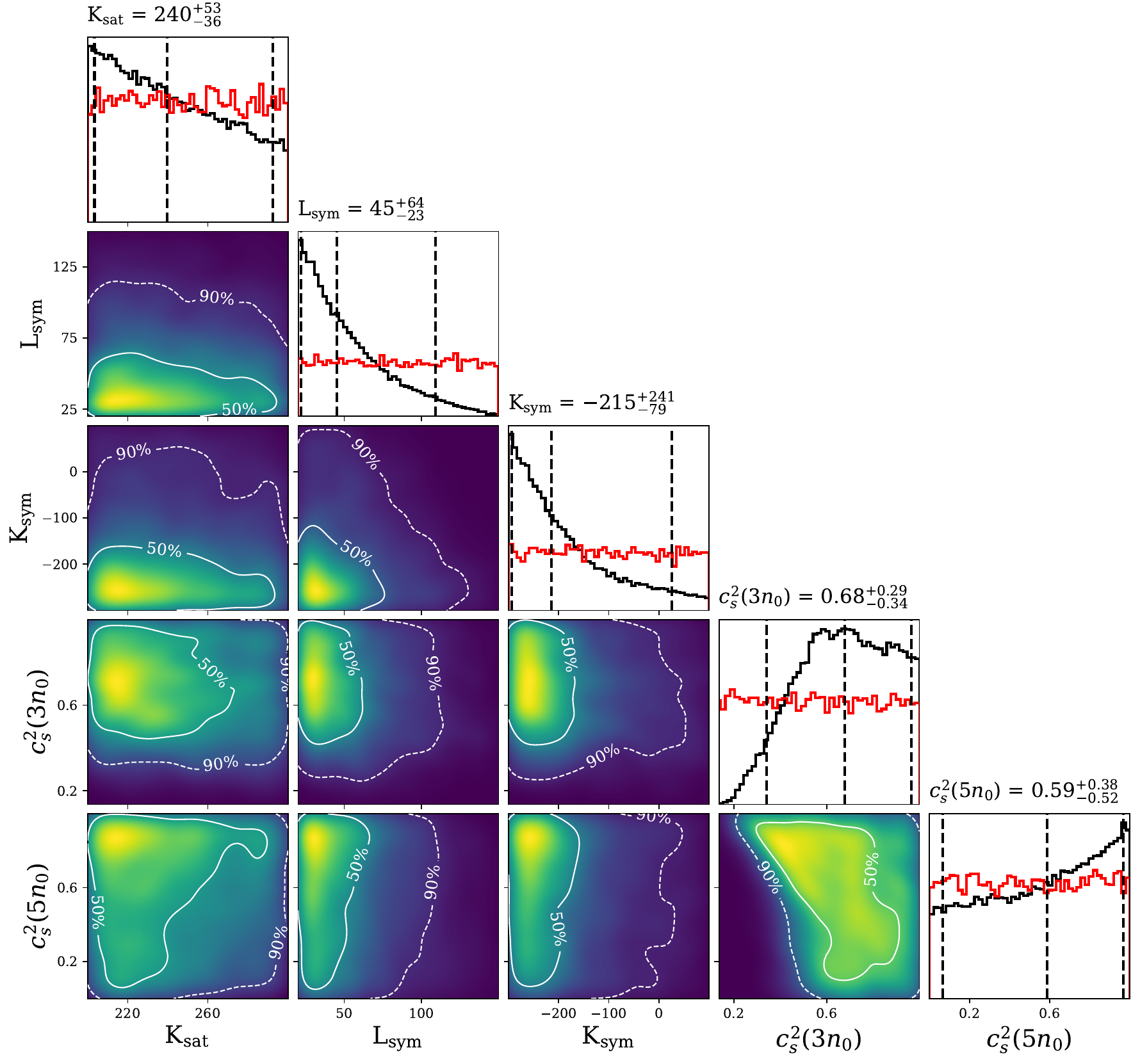}
    \caption{Corner plot of the EOS parameters for the 5-parameter model. 
    We state the central values with their 90\% credible levels.
    The values of the $c_s^2$ are in units of the speed of light squared ($c^2$), while the other EOS parameters are given in units of MeV. 
    We also show the 1D histograms of the prior distribution (red) and posterior from the emulators (black).}
    \label{fig:NEP_corner}
\end{figure}

For the remainder of the results section, we present posteriors from the 8000 live point runs.
Posteriors on the EOS parameters obtained using emulators for the 5-parameter model are shown in \cref{fig:NEP_corner}. 
We find that GW observations provide information on all 5 parameters, with $K_{\rm sat}$, $L_{\rm sym}$, and $K_{\rm sym}$ running against the lower prior bounds.
We constrain the slope and curvature of the symmetry energy at the 90\% upper credible interval to be $L_{\rm sym}\lesssim106$~MeV and $K_{\rm sym}\lesssim26$~MeV, favoring soft \red{low-density} EOSs but with large uncertainties.
The nuclear incompressibility $K_{\rm sat}$ is less well constrained by the data and we find $K_{\rm sat}\lesssim290$.

We also find a preference for large values of $c_s^2(3n_0)$, peaking at values of $c_s^2(3n_0)\sim 0.7$ with values below $c_s^2\sim 0.2$ being ruled out.
This preference implies that our model favors the presence of strongly interacting matter at $3n_0$, which is mostly due to the requirement of the EOS to reproduce the maximum NS mass.
At $5n_0$, however, the sound speed is largely unconstrained by the data because the neutron stars in GW170817 have central densities lower than $5n_0$, consistent with nuclear models showing that canonical neutron stars are most sensitive to the EOS at $2-3n_0$~\cite{2011PhRvC..84f4302F,2023PhRvC.107d5802S,Koehn:2024set}.
However, we find an anti-correlation between these two sound speeds due to requirements for having a sufficiently stiff \red{high-density} EOS to reproduce $2$$M_\odot$ neutron stars and a sufficiently soft \red{low-density} EOS to satisfy the tidal deformabilities favored by GW170817~\cite{Reed:2020}.

\begin{figure}
    \centering
    \includegraphics[width=1\linewidth]{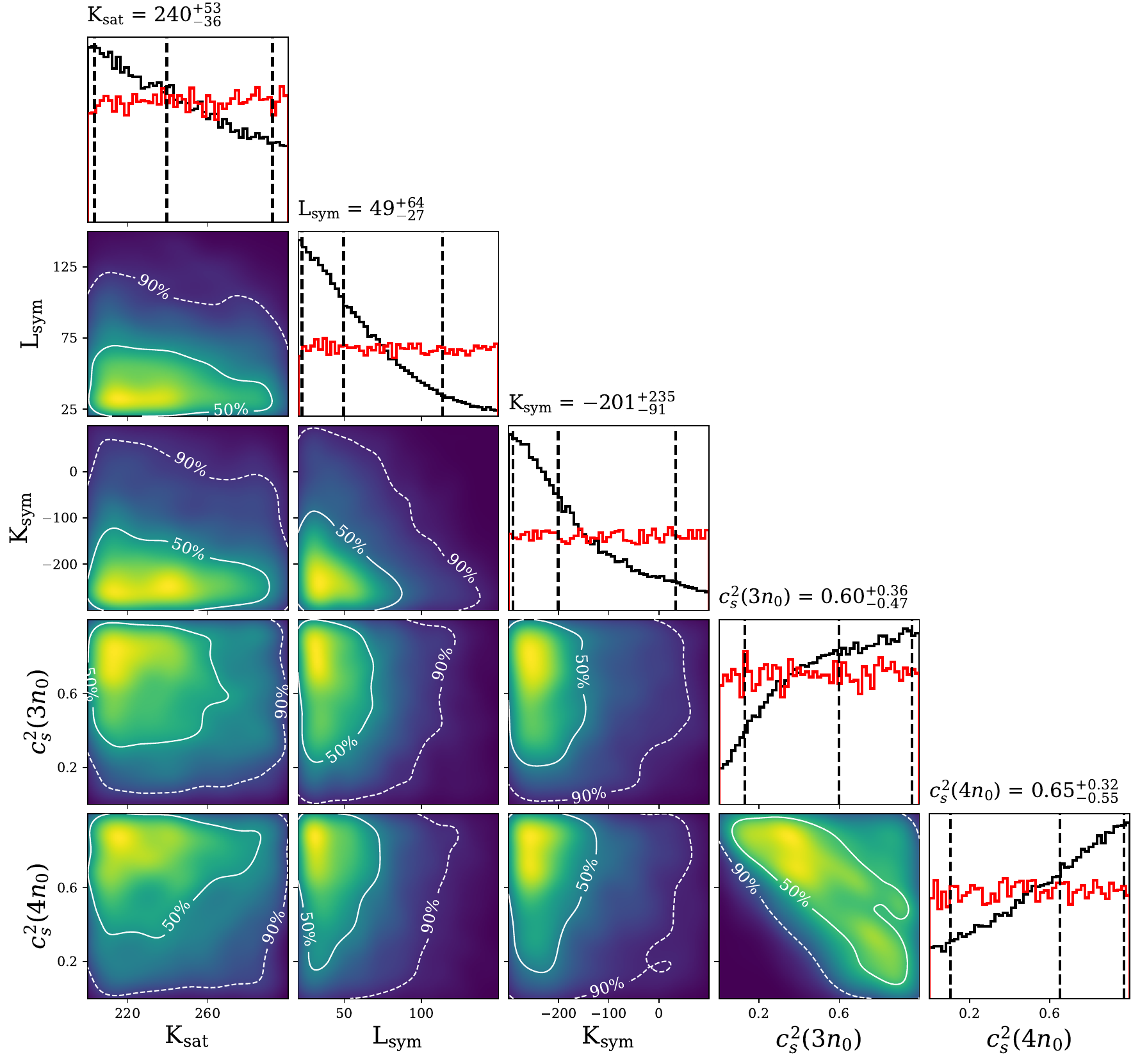}
    \caption{Same as \cref{fig:NEP_corner} but for the 10-parameter model and showing the two lowest-density sound speeds.}
    \label{fig:10par_1}
\end{figure}

\begin{figure}
    \centering
    \includegraphics[width=1\linewidth]{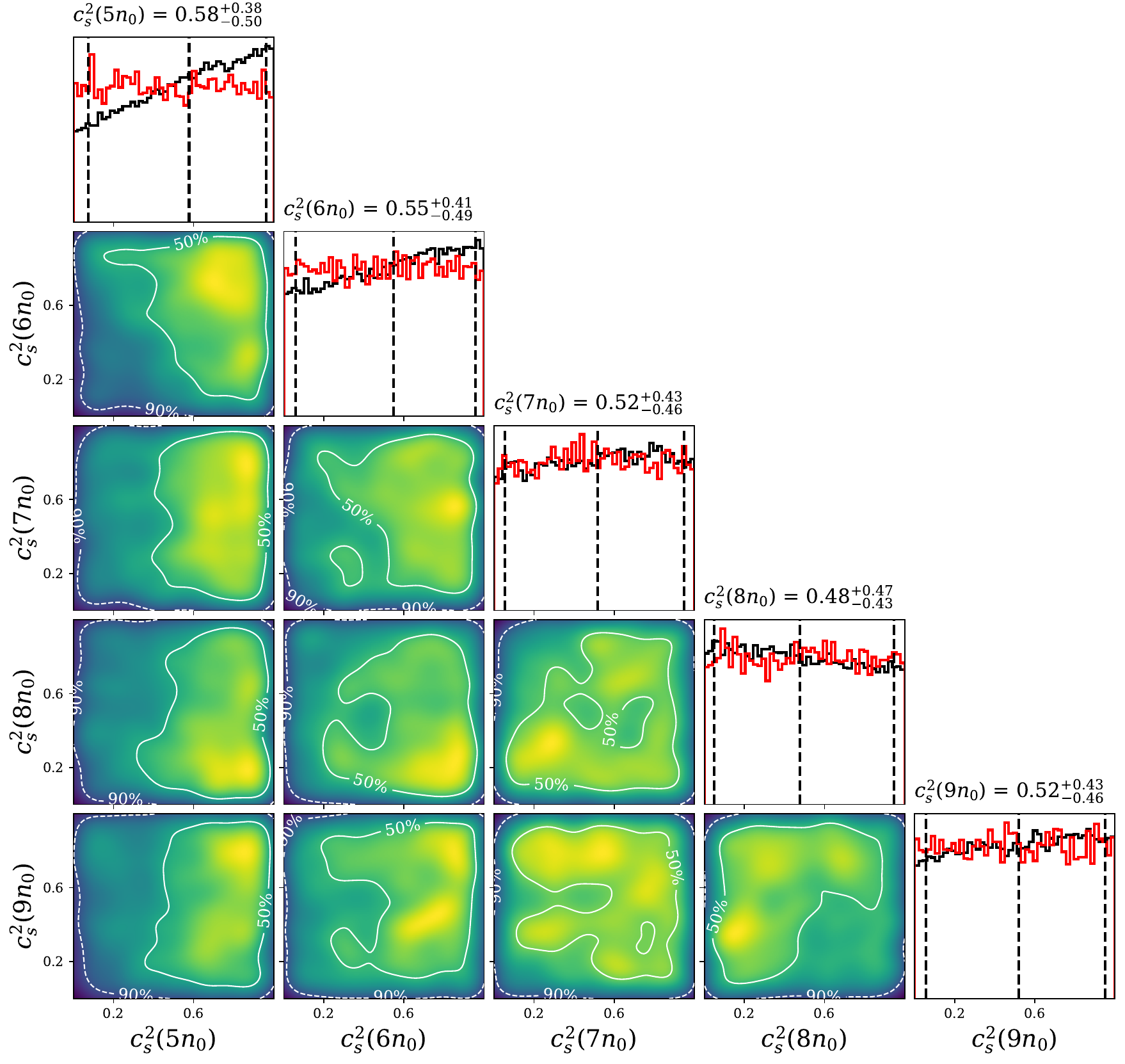}
    \caption{Continuation of \cref{fig:10par_1}, showing the remaining sound-speed parameters of the model.}
    \label{fig:10par_2}
\end{figure}

We show the posteriors on the 10-parameter model in two histograms: \cref{fig:10par_1} shows the three nuclear parameters and the two lowest-density sound speeds and \cref{fig:10par_2} shows the posteriors for the remaining sound speeds. 
The results for the nuclear parameters are similar for the 5- and 10-parameter models, with nearly identical results for $K_{\rm sat}$ and only 5-10\% differences for $L_{\rm sym}$ and $K_{\rm sym}$.
The sound speeds at $3n_0$ and $4n_0$ are again constrained by the data and anti-correlated, similar to the 5-parameter model.
As before, our inference does not constrain the sound speed at larger densities, indicating that the two NSs in GW170817 did not explore this density range.
For this reason, the 5- and 10-parameter models are nearly identical in their complexity when describing GW170817, and in the following we will only use the 5-parameter model to calculate all observables.
We conclude that information on the EOS and its microphysics cannot be directly inferred from GW170817 above $\approx4-5n_0$.

\subsection{Indirect Inference Constraints}

\begin{figure}
    \centering
    \includegraphics[width=1\linewidth]{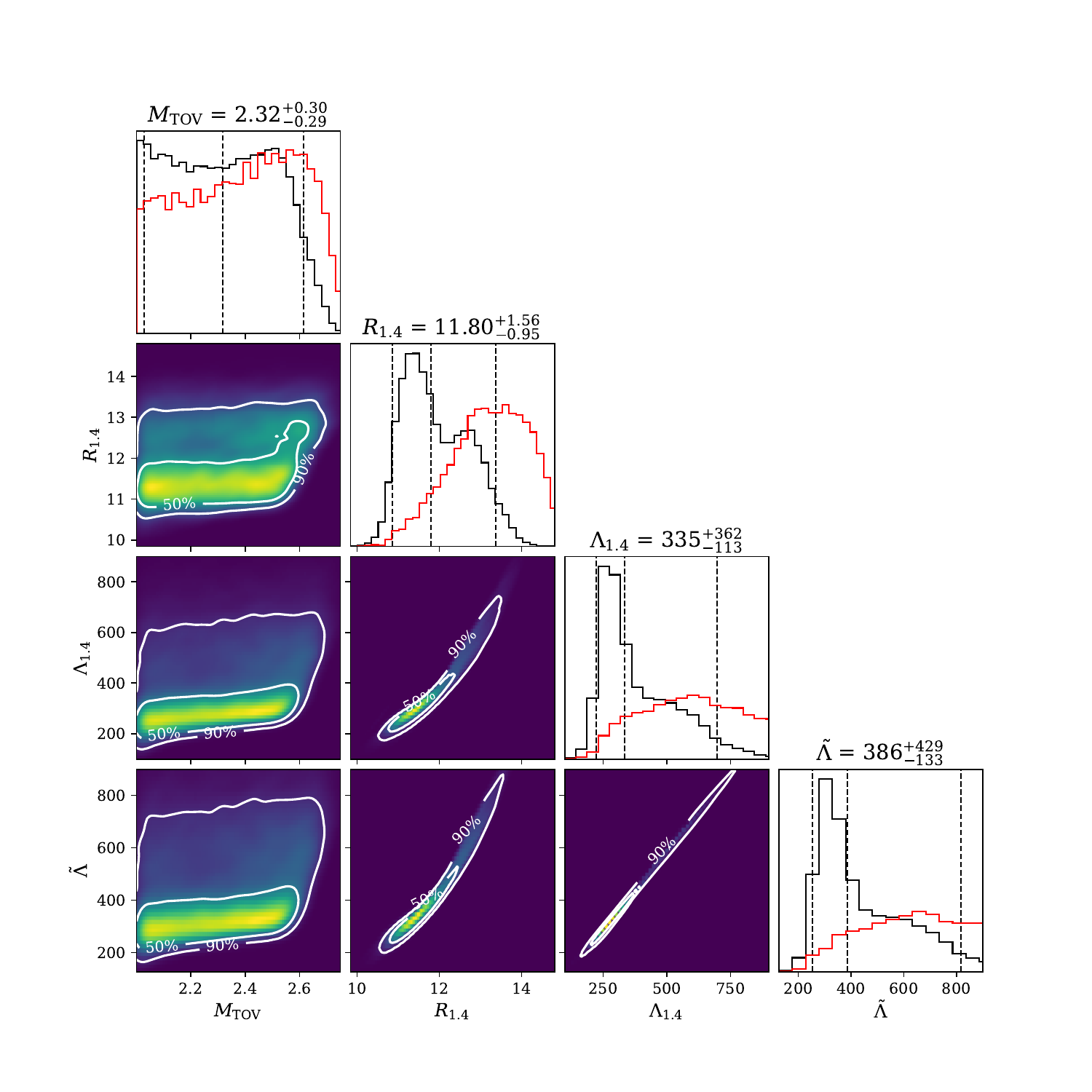}
    \caption{Corner plot of various NS observables using the posteriors for the 5-parameter EOS model.
    We state the central values with 90\% credible intervals and show the 1D marginalized histograms for the prior (red) and the posterior (black).
    The maximum NS mass $M_{\rm TOV}$ is given in units of $M_{\odot}$ and R$_{1.4}$ in units of km.}
    \label{fig:NS_obs}
\end{figure}

With the posteriors on the EOS parameters, we are now able to provide indirect constraints on several NS structure properties.
For this, we propagate the posteriors of \cref{fig:NEP_corner} to all relevant NS observables not calculated by our MLP emulators using our full TOV solver.
We show these observables in \cref{fig:NS_obs}.

\begin{figure}[t]
    \centering
    \includegraphics[width=.5\linewidth]{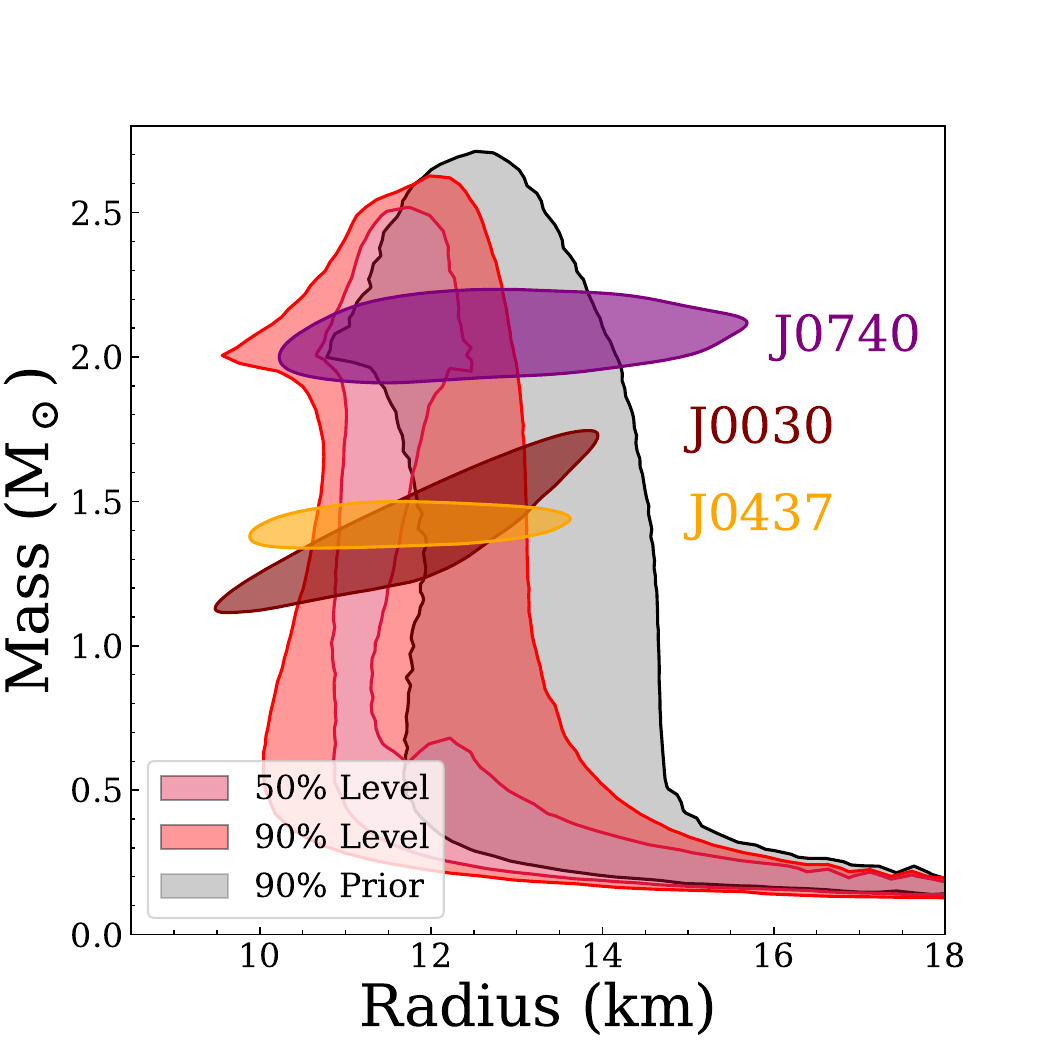}
    \caption{The posterior of the mass-radius relation  for the 5-parameter EOS model using the emulator inference run. 
    We give the 50\% and 90\% credible regions
    in light and dark red, respectively.
    We also show the 90\% credible interval of the prior obtained from the uniform distribution of 10,000 NEP configurations in gray. 
    In addition, we show  posteriors on masses and radii obtained by the NICER collaboration for pulsars J0030 (maroon) \cite{Riley:2019yda,Raaijmakers:2019qny,Vinciguerra:2023qxq}, J0437 (orange) \cite{Choudhury:2024} and J0740 (purple) \cite{Raaijmakers:2021,Riley:2021pdl} at the 95\% credible level.}
    \label{fig:MR}
\end{figure}

We find a maximum NS mass $M_{\rm TOV}$ of $2.32\pm0.21\ M_\odot$ which is consistent with observations of heavy pulsars \cite{Riley:2021pdl,Fonseca:2021wxt} \red{and estimates from other studies \cite{Cai:2025nxn}.}
Our results are slightly biased towards heavy masses $\gtrsim2.4$\msun due to the prior. 
Our posteriors, however, indicate that very large maximum masses are disfavored by GW170817, with a sharp drop-off at $\approx2.5M_\odot$.

We also constrain the radius of a canonical $1.4M_\odot$ NS to be $11.8^{+1.1}_{-0.7}$~km. 
While this result may be somewhat dependent on the crust we use here \cite{Newton:2011dw}, the radius uncertainty is mainly related to the uncertainties on $L_{\rm sym}$ and $K_{\rm sym}$ which contribute considerably to the pressure of neutron-star matter, and hence, to the NS radius.
We find that GW170817 provides strong information on the radius: our prior favors EOS with larger radii of $\approx14$ km and our inference decreases radii by about 2 km.
Similarly, we find the tidal deformability of a $1.4M_\odot$ NS to be $\approx335_{-113}^{+362}$ at the 90\% level, which is reduced from its prior maximum of $\sim650$.
This value is consistent with the original 90\% upper limit constraint from GW170817 of $\Lambda_{1.4}\le800$ \cite{LIGOScientific:2017vwq} and follow-up constraints using generic equation of state models~\cite{LIGOScientific:2018cki}. 
We find the combined deformability of GW170817 $\tilde{\Lambda}$ to be  $\approx386_{-133}^{+429}$. 
In \cref{fig:MR}, we show the full mass-radius posterior we obtain from the inference using the 5-parameter EOS model.
We find that our inferred radii are consistent with observations by NASA's Neutron Star Interior Composition Explorer (NICER) mission \cite{Riley:2021pdl,Riley:2019yda,Miller:2019cac,Raaijmakers:2019qny,Raaijmakers:2021,Miller:2021qha,Choudhury:2024} at the 95\% credible level.

\section{Discussion}
\label{sec:discussion}

First, we highlight the efficiency and accuracy of using emulators in GW inference studies.
This study is using a relatively straightforward inference calculation wherein many of the waveform parameters were tightly constrained (e.g. $\mathcal{M}_c$) or fixed entirely (e.g. sky location).
In a GW inference for a newly detected signal, these sampling parameters are necessary to fully characterize the signal.
This would significantly increase the number of likelihood function calls and slow down the inference.
To simultaneously infer both GW \textit{and} EOS parameters for such a signal would further complicate the matter.
Through the use of emulators, we gain a significant computational advantage over using the full solver when performing such calculations.

Using a traditional on-the-fly TOV solver alongside the full GW inference requires substantially more calculations, and hence, computational time per likelihood evaluation.
In our study, using the full TOV solver required both modeling the EOS \textit{and} solving the NS structure equations.
As stated before, the average time to evaluate the likelihood function using the full solver was $\mathcal{O}(3\rm s)$ which was achieved through careful optimization of the TOV code and data structuring.
In such an approach, a maximum-mass constraint can only by implemented \textit{after} solving for the full $M-R-\Lambda$ sequence, further increasing the computational time.
In comparison, the inferences using emulators bypass all of the steps which is $\approx80$ times faster per likelihood evaluation.
Without these emulators implemented in the Bayesian inference architecture, large-scale inferences of nuclear EOS parameters from GW observations would not be feasible.

We have shown that we can extract meaningful constraints on EOS parameters such as $L_{\rm sym}$ and $K_{\rm sym}$ directly from  GW170817.
These parameters play an important role for neutron-star structure as they constrain the pressure in neutron-rich matter.
For example, the slope of the symmetry energy $L_{\rm sym}$ is related to the pressure in pure neutron matter at saturation density, $P_{\rm PNM}(n_0)=\frac{1}{3}n_0L_{\rm sym}$.
In the absence of phase transitions, NS observables such as radii or tidal deformabilities are correlated to the neutron-matter pressure.
Our constraints on these EOS parameters span a range which has been probed in a variety of different experimental and theoretical calculations.
On the theory side, microscopic calculations~\cite{gandolfi:2010,gandolfi:2012} or density-functional constrains~\cite{Horowitz:2001,FSUGold,dutra:2012} for the nuclear symmetry energy have been available for some time, albeit without robust theoretical uncertainty estimates. 
More recent calculations in the framework of chiral EFT suggest $L_{\rm sym}\sim40-60$~MeV~\cite{Drischler:2017wtt,Drischler:2020hwi,Somasundaram:2020chb,Tews:2024owl} and $K_{\rm sym}$ near $\sim-150$ MeV~\cite{Somasundaram:2020chb}.
Experimentally, the most stringent constraints on the density dependence of the symmetry energy come from measurements of isovector nuclear properties.
The electric dipole polarizability $\alpha_D$ in neutron-rich nuclei \cite{PbAlphaD,NiAlphaD,SnAlphaD} has been shown to have a robust isovector signature which constrains $L_{\rm sym}$ \cite{Piekarewicz:2010fa}.
Experiments measuring $\alpha_D$ in neutron-rich nuclei have constrained $L_{\rm sym}\lesssim65$ MeV \cite{Roca-Maza:2015eza,Reinhard:2021,vonNeumann:2025}.
Similarly, parity-violating electron scattering experiments on neutron-rich nuclei \cite{2021PhRvL.126q2502A,2022PhRvL.129d2501A} have led to calculations of the neutron-skin in these nuclei, a quantity which has been shown to be strongly linked to the slope and curvature parameters of the symmetry energy \cite{Reed:2021nqk,Reed:2024c}.
The PREX experiment led to a constraint of $L_{\rm sym}=106\pm37$ MeV \cite{Reed:2021nqk,Reinhard:2021}, which is consistent with our upper limit but with large uncertainties.
\red{Additionally, constraints from analyses of many astrophysical sources find similar results \cite{Li:2021thg}.}
Implementing constraints from both theory and experiment in a future study may produce stricter constraints than  presented here by employing correlations between nuclear EOS parameters, nuclear structure properties, and astrophysical observations of neutron stars \cite{Koehn:2024set}.

In this work, we have employed a generic EOS model that enables us to account for the appearance of new degrees of freedom at higher densities.
Our results are biased by this choice of EOS model and other choices may produce slightly different results.
However, the framework developed here is flexible and can easily be adapted for studies with other parametric EOS models. 
For example, we can use this framework to study traditional density functional theories~\cite{giuliani2023bayes,Patra:2025}, or models incorporating explicit choices of exotic degrees of freedom, such as hyperons~\cite{Kochankovski:2025} or quarkyonic matter~\cite{Pang:2023dqj}, potentially constraining these degrees of freedom directly from GW observations. 

With the growing need for big data analyses in astrophysics and nuclear physics, the computational resources to process and provide valuable scientific information need not only a lot of time but also consume a considerable amount of energy.
The speedup obtained via the emulators as shown in \cref{tab:cpu_hours}, at an average power consumption per node on NERSC of 100W \cite{10.1145/3624062.3624200}, results in saving 2.2 kWh of energy.
With multiple runs and follow-up studies necessary for a full GW inference, this energy savings add up substantially.
Therefore, utilizing efficient and accurate emulators can potentially have an important environmental and financial impact for science.

\section{Conclusion}
\label{sec:conclusion}
We have implemented emulators for the TOV equations based on multilayer perceptrons in the \texttt{PyCBC Inference} framework to directly sample EOS parameters when analyzing GW data from binary neutron star mergers. 
We have used these tools to perform a Bayesian inference study on the GW data observed from the binary neutron star event GW170817 and constrained EOS parameters for two different EOS models. 
We have found that GW170817 constrains the nuclear equation of state up to approximately 4 times nuclear saturation density, allowing us to extract the corresponding parameters.

In addition to the nuclear EOS, we also constrain the maximum NS mass, radius, and tidal deformability of a $1.4M_\odot$ NS. 
Our study is consistent with measurements of NS radii from the NICER telescope as well as earlier studies of GW170817. 
Our work shows that emulators can considerably enhance the speed of large-scale Bayesian inference studies. 
The framework presented here can also be easily transferred to other parametric EOS models such as models from density functional theory and effective field theories.

\ack
We would like to thank Pablo Giuliani, Marc Salinas, Steven P. Harris, Alexander Nitz, and Chuck Horowitz for helpful discussions related to this work. 
This work also benefited from discussions at the Third Frontiers in Nuclear Astrophysics Summer School at Ohio University, supported by IReNA under National Science Foundation Grant No. OISE-1927130.

B.T.R. was supported by the Laboratory Directed Research and Development (LDRD) program of Los Alamos National Laboratory (LANL) under project number 20230785PRD1.
B.T.R., C.L.A., R.S., S.D., and I.T. were supported by the LDRD program of LANL under project number 20230315ER.
B.T.R. and C.L.A. were also supported by LANL through its Center for Space and Earth Science, which is funded by LANL’s LDRD program under project number 20240477CR.
I.T. was also supported by the U.S. Department of Energy through LANL. 
LANL is operated by Triad National Security, LLC, for the National Nuclear Security Administration of U.S. Department of Energy (Contract No.~89233218CNA000001).
R.S. and I.T. were also supported by the U.S. Department of Energy, Office of Science, Office of Advanced Scientific Computing Research, Scientific Discovery through Advanced Computing (SciDAC) NUCLEI program. 
C.D.C. acknowledges support from NSF Grants No.~PHY-2412341 and AST-2407454.

This research used resources of the National Energy Research Scientific Computing Center (NERSC), a Department of Energy User Facility using NERSC award NP-ERCAP 32335.

This research has made use of data or software obtained from the Gravitational Wave Open Science Center (gwosc.org), a service of the LIGO Scientific Collaboration, the Virgo Collaboration, and KAGRA. 
This material is based upon work supported by NSF's LIGO Laboratory which is a major facility fully funded by the National Science Foundation, as well as the Science and Technology Facilities Council (STFC) of the United Kingdom, the Max-Planck-Society (MPS), and the State of Niedersachsen/Germany for support of the construction of Advanced LIGO and construction and operation of the GEO600 detector. 
Additional support for Advanced LIGO was provided by the Australian Research Council. 
Virgo is funded through the European Gravitational Observatory (EGO), by the French Centre National de Recherche Scientifique (CNRS), the Italian Istituto Nazionale di Fisica Nucleare (INFN) and the Dutch Nikhef, with contributions by institutions from Belgium, Germany, Greece, Hungary, Ireland, Japan, Monaco, Poland, Portugal, Spain. 
KAGRA is supported by Ministry of Education, Culture, Sports, Science and Technology (MEXT), Japan Society for the Promotion of Science (JSPS) in Japan; National Research Foundation (NRF) and Ministry of Science and ICT (MSIT) in Korea; Academia Sinica (AS) and National Science and Technology Council (NSTC) in Taiwan.

\section*{References}

\bibliography{bib}
\bibliographystyle{bibstyle}

\end{document}